\documentstyle[12pt,aasms4]{article}
\begin{document}
\lefthead{Chen \& Huang}
\righthead{Superwind Bow Shock Model in NGC 4945}

\newcommand{\gsim}{\mbox{\hspace{.2em}\raisebox{.5ex}{$>$}\hspace{-.8em}\raisebox{-.5ex}{$\sim$}\hspace{.2em}}}
\newcommand{\lsim}{\mbox{\hspace{.2em}\raisebox{.5ex}{$<$}\hspace{-.8em}\raisebox{-.5ex}{$\sim$}\hspace{.2em}}}
\newcommand{\beq}{\begin{equation}}	\newcommand{\eeq}{\end{equation}}
\newcommand{\bey}{\begin{eqnarray}}	\newcommand{\eey}{\end{eqnarray}}
\newcommand{\etal}{{\em et al.\/}}
\newcommand{\ie}{{\em i.e.\/}}	\newcommand{\eg}{{\em e.g.\/}}
\newcommand{\lt}{\left}	\newcommand{\rt}{\right}
\newcommand{\ssst}{\scriptscriptstyle}
\newcommand{\E}[1]{\times 10^{#1}}

\newcommand{\s}{\,{\rm s}} 	
\newcommand{\yr}{\,{\rm yr}}	\newcommand{\Hz}{\,{\rm Hz}}
\newcommand{\cm}{\,{\rm cm}}	\newcommand{\km}{\,{\rm km}}
\newcommand{\kmps}{\km\s^{-1}}
\newcommand{\parsec}{\,{\rm pc}}\newcommand{\Mpc}{\,{\rm Mpc}}
\newcommand{\um}{\,\mu{\rm m}}
\newcommand{\ergs}{\,{\rm ergs}}	\newcommand{\K}{\,{\rm K}}
\newcommand{\eV}{\,{\rm eV}}	

\newcommand{\za}{z_{1}}		\newcommand{\zb}{z_{2}}
\newcommand{\zo}{z_{0}}		\newcommand{\tha}{\theta_{app}}
\newcommand{\va}{v_{1}}		\newcommand{\vb}{v_{2}}
\newcommand{\Ys}{Y_{\ast}}
\newcommand{\cs}{\xi_{\ast}}	\newcommand{\ts}{\theta_{\ast}}
\newcommand{\Dy}{\Delta_{Y}}	\newcommand{\Dv}{\Delta_{v}}
\newcommand{\ro}{r_{\ssst 0}}
\newcommand{\Ha}{\mbox{H$\alpha$}}

\title{THE CONE-LIKE \Ha\ NEBULA IN NGC~4945: A GALACTIC SUPERWIND BOW SHOCK ?}

\author{Yang Chen \& Jie-Hao Huang}
\affil{Department of Astronomy, Nanjing University, Nanjing 210093, China\\
E-mail: ygchen, jhh@nju.edu.cn}

\begin{abstract}

We find that a non-axisymmetric bow shock model, with an appropriate choice
of parameters, could fit the line splitting velocity field of the cone-like 
\Ha\ nebula in NGC~4945 better than a canonical cone model.
Meanwhile, the bow shock model could also reproduce the morphology of the
\Ha\ nebula.
The bow shock results from the interaction of the galactic superwinds with
a giant HII region.
It is implied that the starburst ring or disk around the galactic nucleus
should be generating strong winds, 
and the bright \Ha\ knot northwest of the nucleus be suffering an anisotropic
mass loss process.

\keywords{galaxies: starburst --- galaxies: individual: NGC~4945 ---
galaxies: kinematics and dynamics --- shock waves}

\end{abstract}
\section{Introduction}

It is now believed that the galactic superwinds may arise from the Seyfert,
AGN, and starburst galaxies, and are able to create a giant bipolar wind
cavity toward the galactic halo.
The complicate phenomena in the cavity region are closely related to the
nature of the superwinds and the galaxy itself.
The edge-on spiral galaxy NGC~4945 has a wind cavity, and the observation
of the split lines has been made (Heckman, Armus, \& Miley 1990, hereafter
HAM).
The P.A.$=45^{\circ}$ line splitting region is $\sim500\parsec$ across, and
the P.A.$=135^{\circ}$ split lines are only found to begin from a distance
$\lsim70\parsec$ to the galactic nucleus.
Recently the conic wind cavity is displayed clearly in optical and near
infrared wavelengths (Moorwood \& Oliva 1994; Moorwood \etal\ 1996,
hereafter MVKMO).
The \Ha\ image shows a conic nebula of an opening angle $78^{\circ}$ inside
an infrared fan-shaped structure.
The nucleus is invisible in \Ha, but is coincident with the position of the
peaked emission of the $K$ band and the Br$\gamma$ line.
The spectrum of the head portion of the \Ha\ nebula is typical of LINERs.

It is suggested (MVKMO) that the cone-like \Ha\ nebula can be explained with
Suchkov \etal's (1994, hereafter SBHL) theory on superwind bubble, 
in which the biconic cavity wall begins from a nozzle, and then stretches out
like a cone.
It could, however, be found difficult to fit the line splitting velocity
field data with such a cone.
For the position of the P.A.$=45^{\circ}$ slit on the nebula, the cavity wall
could be represented, as an example, by a cone described by HAM in concept,
in which the materials flow out (at the speed $V_{0}$) along a biconic
surface centered on the nucleus, with the cone's symmetry axis along the
minor axis of the galaxy.
The origin of the coordinates is set at the central nucleus, and the $z$
axis is set along the minor axis, with the positive position toward the
northwest.
The aspect angle, namely the inclination of the line of sight to the galactic
plane, is denoted as $\phi$.
If a distance $d=4.6\Mpc$ is adopted for NGC~4945 (HAM), and the ordinates
of the cutting point of the $z$ axis by the slit for the P.A.$=45^{\circ}$
line splitting is written as $(0,0,\zo)$, 
then the projected distance of the cutting point to the nucleus is
$\zo\cos\phi$, measured $300\parsec$.
With the parameters suggested by HAM, $\phi=33^{\circ}$ and $V_{0}=525\kmps$,
the fitting result is plotted as dotted lines in Fig.1.
The curve for P.A.$=45^{\circ}$ is of a feature something like a polished
triangle, with a peaked top and a flattened bottom.
This typical tendency will not be changed appreciably, if different values
of parameters are adopted, and even if an expansion velocity component
perpendicular to the cone surface is allowed.
Though, the scenario that materials flow along a cone-like surface is
constructive and indicative of another mechanism proposed as follows.

On the other hand, as seen in the \Ha\ image of the nuclear region of
NGC~4945 (Fig.1d \& Fig.3a in MVKMO),
there is not a visual ``nozzle'' as described by SBHL for this cone-like
nebula,
and the apex of this nebula is not located at the central nucleus, but
rather nearly on the linking line between the nucleus and the bright
\Ha\ knot at the position ``C'' denoted there.
The head portion of the nebula surrounds the bright \Ha\ knot.
This clear-cut cometary \Ha\ nebula with a noticeable position of the
\Ha\ knot closely behind the nebular apex seems more likely to be a bow
shock structure than a superwind cone.
We find that not only the presence of a bow shock is possible because of
the interaction of the galactic superwind with the \Ha\ knot,
but also the bow shock model could reproduce the morphology of the
\Ha\ nebula and fit the line splitting velocity field well.

\section{The bow shock model}

As we can see from the \Ha\ image of the nuclear region of NGC~4945 (MVKMO),
the bright \Ha\ knot at ``C'' (denoted by MVKMO) has a nucleus-halo structure.
And as HAM pointed out, the bright \Ha\ knot at ``C'' is evidently a giant
HII region, as well as other knots nearby.
Based on the observation of the HII regions and the accordingly established
Champagne model for them (Yorke 1986; Osterbrock 1989),
the HII region with a nucleus-halo structure should undergo a star formation
activity (\eg, Ye 1992),
and the stellar winds from the O stars in this type of HII region is
undoubtedly important (see, \eg, Osterbrock 1989).
We would assume that the kinetic energies of the winds of the individual O
stars could be released out easily in spite of the collisions among themselves.
Shock heating related to the ongoing starburst would be a dominant excitation
mechanism for this region (see, \eg, Dopita \& Sutherland 1995; Cozial 1996;
Ho, Fillippenko, \& Sargent 1993),
and could be responsible for the LINER-like spectrum of region ``C''
obtained (MVKMO).
According to Moorwood \& Oliva (1994), the superwind of NGC~4945 is driven
by starburst supernovae and/or AGN.
The collision of the galactic superwind with the stellar winds from the HII
region at ``C'' would inevitably produce a bow shock (\eg, Huang \& Weigert
1982; Bandiera 1993; etc.), 
and hence could account for the cone-like \Ha\ nebula.
\Ha\ cometary nebulae have been very common in the stellar level and
attributed to bow shocks, \eg\ PSR~1957+20 (Kulkarni \& Hester 1988), NGC~2359
(Schneps \& Wright 1980; Chen, Wang, \& Qu 1995), etc., 
and the nebula in NGC~4945 seems very likely to be a bow shock phenomenon
in the galactic level.
In the following we would simulate the line splitting velocity field and
the morphology of this nebula using a bow shock model.

From MVKMO's image, the projected distances of the apex of the \Ha\ nebula
to the knot at ``C'' and the nucleus measure $\ro=53\parsec$ and
$z_{a}=66\parsec$, respectively.
If the superwinds arise purely from the nucleus, the ratio of these two
distances $\sim0.8$ would imply a much bigger opening angle
($\sim160^{\circ}$) for the bow shock (Cant\'{o}, Raga, \& Wilkin 1996).
It should not be the real case, however.
The Br$\gamma$ and $L$ band images reveal a $\sim400\parsec$ size starburst
ring or disk on the galactic plane (Moorwood \& Oliva 1994; MVKMO).
Since this size is comparable with that of the line splitting region of the
\Ha\ nebula, 
the ring/disk can contribute a component of starburst driven winds, and then
it could be expected to alter the whole effect of the superwinds.
Such a wind complex could constrict the bow shock to a smaller opening angle
than the pure nuclear wind, though its momentum flux distribution is
complicated.
It is assumed for simplicity that the sum of the two components of winds can
be approximated as a wind with an average velocity $\va$ which acts as if it
is generated from a convergence point $\za$ ($<0$) far behind of the nucleus.

As can be seen carefully in the \Ha+[NII] image, the head portion of the
nebula is not axisymmetric with respect to the axis linking the nucleus and
the knot at ``C'':
the nebular surface at the east is close to the knot and even a slight dent
can be discerned, while the surface at the south is protrusive.
This morphological dent could be due to the relatively low momentum flux of
the giant HII region's wind on its equatorial plane.
When its equatorial plane is tilted, the asymmetry appears.
Therefore, for the giant HII region, we would use a bipolar momentum flux
distribution for the wind, which has been widely used in the stellar level.  
Furthermore, as the equatorial plane is tilted to the left, the left- and
right-hand sides of the bow shock shell would swell up, and relatively
the back- and front-sides would shrink inward (cf.\ the morphology simulation
below).
Consequently, the polished triangle configuration in the fitted velocity
field for the axisymmetric cone would be changed, 
and especially the top part would be expected not to be peaked due to the
flattened back side of the shell.
An algorithm for calculating the asymmetric bow shock has been developed by
Bandiera (1993), and would be adopted in our simulation as follows.

The subscript 1 would refer to the quantities of the superwind complex, and
2 to the quantities of the wind from the \Ha\ knot.
The momentum loss per steradian from the \Ha\ knot is presumably described
by (Bandiera 1993)
\[ Y_{2}=\Ys[1+\Dy(1-3\cos^{2}\chi)/2]. \]
Here $\Ys$ is the average momentum loss per steradian.
$\Dy$ (at a range $-2\le\Dy\le1$) represents the amount of anisotropy in the
wind momentum flux; when $\Dy<0$, $Y_{2}$ tends to be a bipolar distribution.
And $\chi$ is the angle between a given direction and the wind's axis of
symmetry.
The direction of the symmetry axis is represented by the azimuthal and polar
angles $(\ts,\cs)$ with reference to the $z$ axis.

For the sake of making the freedom degree of the used parameters the least,
the superwind complex is regarded as isotropic in the nebular region, 
and the velocity $\vb$ of the wind from the giant HII region as independent
of directions first, 
which is presumably typified by the FWHM $\sim600\kmps$ of the broad feature
in the spectrum of region ``C'' (MVKMO).
Now the free quantities used to fit the data of the line splitting and the
nebular morphology are $\za$, $\va$, $\ts$, $\cs$, $\Dy$, and $\phi$.
Each of these quantities could sensitively affect some certain features in
the outcome.
Basically, $\za$ affects the spatial range of the P.A.$=45^{\circ}$ split
lines;
$\va$ the blueshift and redshift values for the split lines;
$\cs$ the projected position for the slight dented distortion in the
nebular head portion;
$\ts$ the projected position for the slight distortion in the head and the
inclination of the closed curve for the line splitting velocity field,
$\Dy$ the distortion degree of the head and the shape of the closed curve,
and 
$\phi$ the height of the fitting curves for the shifted velocities.
Therefore there is actually very little freedom in the determination of the
parameters, though the determination may not be unique.

To calculate the blue- and redshift of the velocities, we project the
streaming velocity along the bow shock surface to the line of sight.
By repeated tests and comparisons, some parameters could be decided first:
$\za-z_{a}=-16\ro$, $\ts=180^{\circ}$, $\cs=60^{\circ}$, and
$\phi=14^{\circ}$,
and then a solid and dashed lines are plotted in Fig.1 for $\va=1400\kmps$,
$\Dy=-1.5$ and $\va=1200\kmps$, $\Dy=-2$, respectively.
These two curves could basically trace all the points of data for the split
lines, and are much better than the curve based on the cone model.
A line splitting region $\simeq500\parsec$ at $z_{0}\cos\phi=300\parsec$ for
the P.A.$=45^{\circ}$ data is accordingly reproduced.
The fitting aspect angle $\phi\sim14^{\circ}$ also seems reasonable for
an {\em almost} edge-on galaxy (Moorwood \& Oliva 1994).
And it is in better agreement with that observed for the galaxy ($12^{\circ}$) 
by Nakai (1989), than that estimated by HAM ($33^{\circ}$).
With the HAM's estimation of the mass loss rate of the nuclear region of
NGC~4945, $1.6M_{\odot}\yr^{-1}$, 
the fitting velocity $\va\sim1400$---$1200\kmps$ would lead to a momentum
flux of $\sim1.4$---$1.2\times 10^{34}\,{\rm dynes}$.
This is in a very good agreement with HAM's estimation of the momentum flux
$\sim$1.3, 2, or 5$\times 10^{34}\,{\rm dyne}\s^{-1}$ in different ways.
Major shortcomings of the above simulation may be that 
1) there is a small bump in both the fitting curves for P.A.$=135^{\circ}$,
and 
2) the two curves for P.A.$=45^{\circ}$ could not reproduce the trend of a
little bit larger splitting of the southwestern half.
They could be understood if we consider that the real environment of the
winds are more sophisticated than described in our relatively simple model.
For example, the small bumps in the fitting curves for P.A.$=135^{\circ}$
could possibly be accounted for with the ignorance of the additional
squeezing effect of the starburst ring's wind to the head portion of the bow
shock in the calculation, 
because this additional effect could make the head portion more acute than
the present simulation.
The $L(3.5\um)$ continuum image (MVKMO) and the bumps in the H$_{2}$
rotation curves seem to be more suggestive of a ring than a disk
(Moorwood \& Oliva 1994).
And the actual bow shock tip might be closer to the nucleus than the
observed apex of the \Ha\ emission, and be heavily extincted; thus the bump
could be expected to be alleviated.
The latter situation would suggest that the wind from the nucleus itself
is not as strong as the wind from the extended part (\ie, the ring/disk).
Considering the asymmetry in the galactic wind complex could possibly
remove the bump and fit the data better, too.
As for the second point, it could be improved by introducing an anisotropic
velocity distribution for the wind from the HII region:
\[ \vb=v_{2a}(1-\Dv\cos\chi), \]
where $\vb$ is assumed to have the same symmetric axis as $\Dy$, $v_{2a}$
is the average velocity of the wind ($\sim600\kmps$), and $\Dv$ represents
the amount of anisotropy in the wind velocity.
While maintaining the parameters adopted above, we take $\Dv=0.5$.
The resulting split velocities are plotted in Fig.2.
The curves for P.A.$=45^{\circ}$ fit the data points and the mentioned trend
pretty well.

The morphology of the bow shock is decided by the momentum flux, and is
independent of the wind velocities.
Taking a set of parameters used above $\za-z_{a}=-16\ro$, $\ts=180^{\circ}$,
$\cs=60^{\circ}$, $\Dy=-1.5$, and $\phi=14^{\circ}$, for instance,
the morphology of the bow shock is demonstrated in Fig.3, which reproduces
the shape of the \Ha\ nebula very well.

In view of the above simulations, an asymmetric bow shock model could indeed
fit both the line splitting data and the nebular morphology.
In the simulation, the galactic gravitation has been ignored.
In fact, one could find that the gravitation effect is not important in the
head portion compared with the wind's ram presure, and that the flow velocities
($\gsim500\kmps$) in the trailing portion is higher than the escape velocity.
The reasons why the stellar winds from the giant HII region could
have anisotropic momentum flux and velocity distributions remain to be
further studied.
Here a preliminary suggestion may be that the HII region is flattened, with
denser materials on its equatorial plane than at the two poles.

\section{Discussion}

The fan-shaped appearance of the filaments emerging from the nucleus of
NGC~4945 discovered by Nakai (1989) is an evidence for the existence of
a superwind activity.
The opening angle of this fan structure ($\sim150^{\circ}$) is much larger
than that of the cone-like \Ha\ nebula ($\sim78^{\circ}$) only within which
the split lines were detected (HAM).
This fact implicates that the filament fan may have a common character with the
infrared fan-shaped structures in $Z$, $J$ bands, in which the extinction is
lower (MVKMO).
So the bow shock \Ha\ nebula is formed within the superwind cavity and is
different in essence from the fan-shaped structures.
The interior of the cone-like \Ha\ nebula should be occupied by the wind
from the HII region, rather than the galactic superwinds.

It is intriguing that, in the \Ha\ image of NGC~4945, there is another
bright \Ha\ knot near the denoted position ``B'' on the opposite side of the
``C'' knot,
aligned symmetrically with respect to the nucleus.
It is also worth noting that the extension of both [FeII] and
H$_{2}$(1--0)$S$(1) line emission are roughly symmetric on both sides of
the galactic plane (as pointed out by Moorwood \& Oliva 1994).
These phenomena of symmetry might have something to do with the
symmetric activity from the nucleus,
most likely a bipolar galactic superwind.
MVKMO suggested that the infrared fan structure is consistent with the
superwind bubble model developed by SBHL.
Indeed, the two bright HII regions discussed above  could naturally be
born as a result of the bipolar superwind.
According to the SBHL model, especially to the B2 model therein, the
superwind entrains the disk matter to the halo.
At very early stage, two dense gaseous blobs, or two clouds, are formed
symmetrically about the galactic nucleus.
Thus two giant HII regions, as we see in the nuclear region of NGC~4945,
might thereafter be hatched in these two dense clouds due to the star
formation activities, giving rise to the stellar winds.

\section{Conclusion}

We have found it difficult to fit the line splitting velocity field of the
conic \Ha\ nebula using a canonical cone; 
with an appropriate choice of parameters, however, it could essentially be
reproduced by a non-axisymmetric bow shock.
The morphology of the bow shock simulated is also very similar to the
\Ha\ nebula.
It is implied that the the starburst ring or disk around the galactic
nucleus should be blowing strong winds.
The bow shock originates from the collision of the superwind complex from
the nucleus and the starburst ring/disk with the northwestern bright
\Ha\ knot, namely with the wind from the giant HII region at ``C''.
The superwind complex could be effectively approximated as a wind at an average
velocity $\sim1400\kmps$ from a convergence point far behind the nucleus.
And the wind from the giant HII region might have anisotropic momentum flux
and velocity distributions.
The best-fit average velocity of the superwind and the aspect angle are in
good agreements with the observational data.

\acknowledgements{We wish to thank Rino Bandiera and Tim Heckman for careful
reading and critical comments for the manuscripts which help to improve our
analyses. We are indebted to R.~B.\ also for providing his elaborate codes
for computing the non-axisymmetric bow shock.
This work was carried out on the SUN workstation at the Laboratory for
Astronomical Data Analysis of the Department of Astronomy, Nanjing University.
This work is supported by a grant from the NSF of China, a grant from the
State Education Commission of China for the scholars back from abroad, and a
grant from the Ascent Project of the State Scientific Commission of China.}

\clearpage

\figcaption[fig1.eps]{The simulation of the line splitting velocity field, the
top figure for P.A.$=45^{\circ}$ and the bottom figure for P.A.$=135^{\circ}$.
The crossed points are adapted from HAM's observational data.
The dotted line is plotted according to the cone model with HAM's parameters 
except that the apparent opening angle $\tha=78^{\circ}$ has been translated
to the real angle $\theta_{rl}$ by projectional effect:
$\theta_{rl}=2\arctan[\tan(\tha/2)\cos\phi]$.
The solid and dashed lines are obtained from the bow shock model for 
$\za=z_{a}-16\ro$, $\va=1400\kmps$, $\ts=180^{\circ}$, $\cs=60^{\circ}$,
$\Dy=-1.5$, $\phi=14^{\circ}$, 
and $\za=z_{a}-16\ro$, $\va=1200\kmps$, $\ts=180^{\circ}$, $\cs=60^{\circ}$,
$\Dy=-2$, $\phi=14^{\circ}$, respectively.
(See the text for the denotation of these quantities.)
\label{fig1}}

\figcaption[fig2.eps]{The simulation of the line splitting velocity field
using the bow shock model with anisotropic distribution for the HII region's
wind velocity.
The top figure is for P.A.$=45^{\circ}$ and the bottom figure for
P.A.$=135^{\circ}$.
The parameters for the solid and dashed lines are the same as in Fig.1,
except that $\Dv=0.5$ in addition for the both cases.
(See the text for the denotation of the quantities.)
\label{fig2}}
\figcaption[fig3.eps]{The projected morphology of the bow shock for
$\za=z_{a}-16\ro$, $\ts=180^{\circ}$, $\cs=60^{\circ}$, $\Dy=-1.5$,
and $\phi=14^{\circ}$.
(See the text for the denotation of these quantities.)
\label{fig3}}

\end{document}